\documentclass[runningheads]{llncs}
\usepackage{cite}
\usepackage{authblk}
\usepackage{subcaption}
\usepackage{amsmath,amssymb,amsfonts}
\usepackage{graphicx}
\usepackage{textcomp}
\usepackage{xcolor}
\usepackage[english]{babel}
\usepackage[utf8]{inputenc}
\usepackage{hyperref}
\usepackage[utf8]{inputenc}
\usepackage[english]{babel}

\usepackage[font={small}]{caption}

\usepackage[linesnumbered,ruled,vlined]{algorithm2e}

\begin{document}
\title{IAD: Indirect Anomalous VMMs Detection in the Cloud-based Environment}
\titlerunning{IAD: Indirect Anomalous VMMs Detection in the Cloud-based Environment}
%
\author{Anshul Jindal\inst{1}\orcidID{0000-0002-7773-5342} \and
Ilya Shakhat\inst{2} \and
Jorge Cardoso\inst{2,3}\orcidID{0000-0001-8992-3466} \and
Michael Gerndt\inst{1}\orcidID{0000-0002-3210-5048} \and
Vladimir Podolskiy\inst{1}\orcidID{0000-0002-2775-3630}}
\authorrunning{Jindal et al.}
%
\institute{Chair of Computer Architecture and Parallel Systems, \\Technical University of Munich, Garching, Germany  \\
\email{{anshul.jindal@tum.de, gerndt@in.tum.de, v.podolskiy@tum.de}}\\ 
\and Huawei Munich Research Center, Huawei Technologies Munich,Germany\\ 
\email{\{ilya.shakhat1, jorge.cardoso\}@huawei.com}
\and
University of Coimbra, CISUC, DEI, Coimbra, Portugal
}
\maketitle              
\begin{abstract}
Server virtualization in the form of virtual machines (VMs) with the use of a hypervisor or a Virtual Machine Monitor (VMM) is an essential part of cloud computing technology to provide infrastructure-as-a-service (IaaS). A fault or an anomaly in the VMM can propagate to the VMs hosted on it and ultimately affect the availability and reliability of the applications running on those VMs.  Therefore, identifying and eventually resolving it quickly is highly important. However, anomalous VMM detection is a challenge in the cloud environment since the user does not have access to the VMM.

This paper addresses this \textit{challenge of anomalous VMM detection in the cloud-based environment without having any knowledge or data from VMM} by introducing a novel machine learning-based algorithm called \textbf{IAD}: \textbf{I}ndirect \textbf{A}nomalous VMMs \textbf{D}etection. This algorithm solely uses the VM's resources utilization data hosted on those VMMs for the anomalous VMMs detection. The developed algorithm's accuracy was tested on four datasets comprising the synthetic and real and compared against four other popular algorithms, which can also be used to the described problem. It was found that the proposed \textit{IAD} algorithm has an average F1-score of 83.7\% averaged across four datasets, and also outperforms other algorithms by an average F1-score of 11\%.
\end{abstract}
\section{Introduction}
\begin{figure}[t]
\centerline{\includegraphics[width=0.5\linewidth]{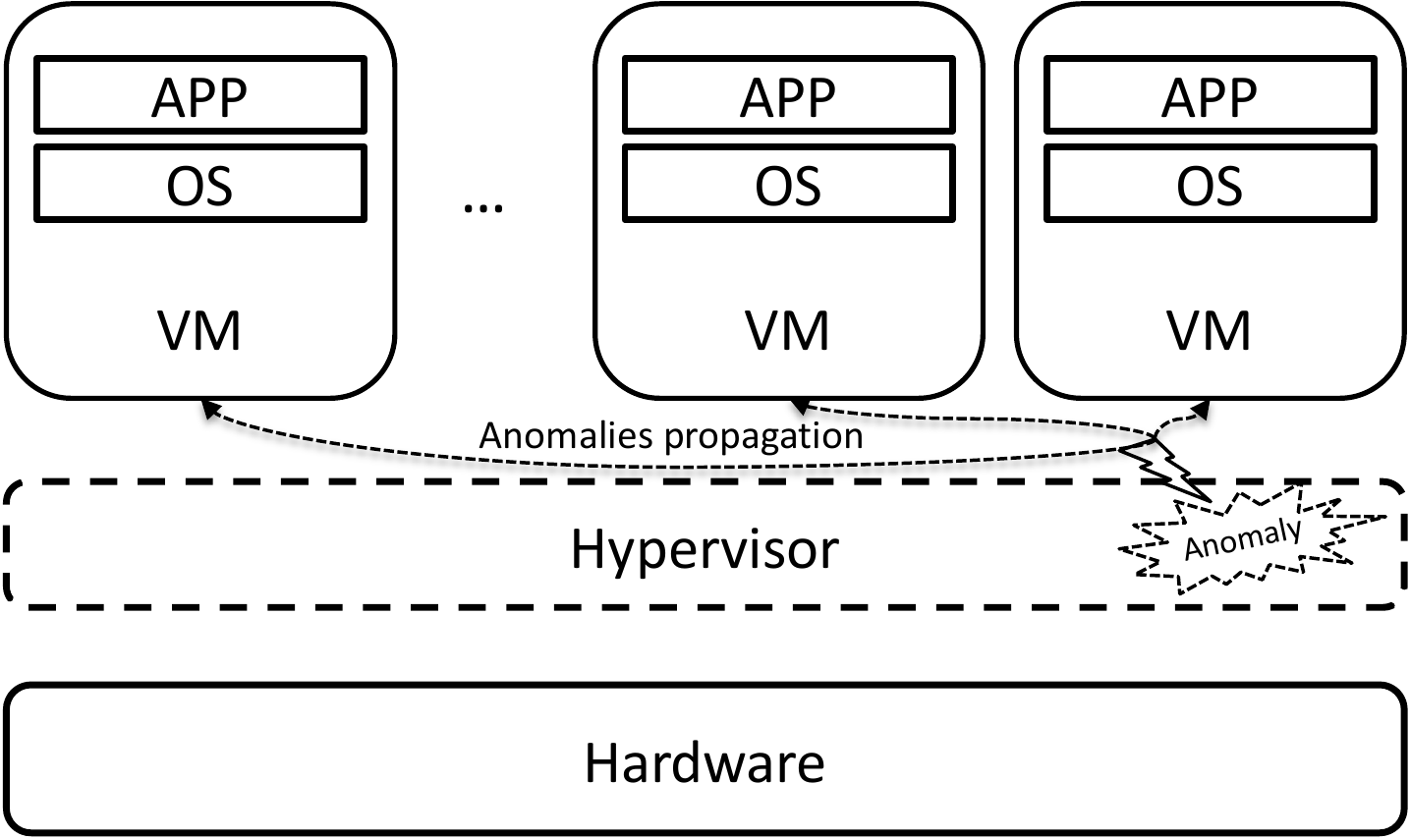}}
\caption{An example showcasing the propagation of anomalies in a Type-1 hypervisor or VMM to the virtual machines (VMs) hosted on it. 
}
\label{fig_example_motivation}
\end{figure}

Cloud computing enables industries to develop and deploy highly available and scalable applications to provide affordable and on-demand access to compute and storage resources. Server virtualization in the form of virtual machines (VMs) is an essential part of cloud computing technology to provide infrastructure-as-a-service (IaaS) with the use of a hypervisor or Virtual Machine Monitor (VMM)~\cite{6530588}. Users can then deploy their applications on these VMs with only the required resources. This allows the efficient usage of the physical hardware and reduces the overall cost. The virtualization layer, especially the hypervisors, is prone to temporary hardware errors caused by manufacturing defects, a sudden increase in CPU utilization caused by some task or disconnection of externally mounted storage devices, etc. The VMs running on these VMMs are then susceptible to errors from the underneath stack, as a result, can impact the performance of the applications running on these VMs~\cite{10.1145/1952682.1952692, Li2008UnderstandingTP}. Figure~\ref{fig_example_motivation} shows an example propagation of anomalies in a virtualization stack using a type-1 hypervisor to the VM hosted on it. These anomalies may lead to the failure of all VMs and, ultimately, the applications hosted on them. 

In the development environment, these anomalous VMMs are relatively easily detectable by analyzing the logs from the hypervisor dumps. But in the production environment running on the cloud, anomalous VMMs detection is a challenge since a cloud user does not have access to the VMMs logs. Additionally, many anomalous VMM detection techniques have been proposed~\cite{6957243, 10.1145/339647.339652, Nikolai2014HypervisorbasedCI}. However, these works either require the monitoring data of the hypervisor or inject custom probes into the hypervisor. Therefore, the usage of such solutions becomes infeasible. Furthermore, due to the low downtime requirements for the applications running on the cloud, detecting such anomalous VMMs and their resolutions is to be done as quickly as possible. 

Therefore, this challenge is addressed in this paper for detecting anomalous VMMs \textit{ by solely using the VM's resources utilization data hosted on those VMMs} by creating a novel algorithm called \textbf{IAD}: \textbf{I}ndirect \textbf{A}nomalous VMMs \textbf{D}etection. We call the algorithm indirect since the detection must be done without any internal knowledge or data from the VMM; it should be solely based on the virtual machine's data hosted on it. The key contributions are : 

\begin{itemize}
    \item We present an online novel machine learning-based algorithm \textbf{IAD} for accurate and efficient detection of anomalous VMMs by solely using the resource's utilization data of the VM's hosted on them as the main metric (\S\ref{sec:iad_algorithm}).
    \item We evaluate the performance of the \textit{IAD} on two different aspects: 1) Anomalous VMMs finding accuracy (\S\ref{sec:est_time_accuracy}), and 2) Anomalous VMMs finding efficiency and scalability (\S\ref{sec:config_finding_efficency_scalability}), and compare it against five other popular algorithms which can also be applied to some extent on the described problem. 
    \item We evaluate the \textit{IAD} algorithm and other five popular algorithms on synthetic and two real datasets.
\end{itemize}
 
\textit{Paper Organization: } Section~\ref{sec:problem_statement} describes the overall problem statement addressed in this paper along with an illustrative example. The design and details of the proposed \textit{IAD} algorithm are presented in Section~\ref{sec:iad_algorithm} . Section~\ref{sec:exp_settings} provides experimental configuration details along with the algorithms and the datasets used in this work for evaluation. In Section~\ref{sec:results}, the evaluation results are presented. 
Finally, Section~\ref{sec:conclusion} concludes the paper and presents an outlook.

\section{Problem Definition}
\label{sec:problem_statement}
This section presents the overall problem definition of indirectly detecting anomalous VMMs in a cloud-based environment. Table~\ref{tab1:symbols} shows the symbols used in this paper.
\begin{table}[t]
\caption{Symbols and definitions.}\label{tab1:symbols}
\begin{tabular*}{\textwidth}{l @{\hskip 0.1in} l}
\hline
\textbf{Symbol} &  \textbf{Interpretation}\\
\hline
$n$ & Number of time ticks in data \\
$d$ & Number of virtual machines hosted on a VMM \\
$X_t$ & The percentage utilization of a resource (for example, CPU  \\
 & or disk usage) by a VM at a time $t$ \\
$X_{t}^j$ & The percentage utilization of a resource at a time $t$ for $j^{th}$ VM  \\

$\{c_{t}^1, c_{t}^2, ..., c_{t}^m\}$ & a set of m $\leq$d VMs with change point at time tick $t$\\

$w$ &  Window size\\
minPercentVMsFault & Minimum \% of total number of VMs on a VMM  which must\\
&   have a change point for classifying the VMM anomalous.  \\

\hline
\end{tabular*}
\end{table}

We are given $X$ = $n \times d$ dataset, with $n$ representing the number of time ticks and $d$ the number of virtual machines hosted on a VMM. $X_{t}^j$ denotes the percentage utilization of a resource (for example, CPU or disk usage) at a time $t$ for $j^{th}$ VM. Our goal is to detect whether the VMM on which the $d$ virtual machines are hosted is anomalous or not. Formally:
\begin{problem}{ (Indirect Anomalous VMM Detection ) }
\begin{itemize}
    \item \textbf{Given} \textit{a multivariate dataset of $n$ time ticks, with $d$ virtual machines ($X_{t}^j$ for $j=\{1,\cdots,d\}$ and $t = \{1,\cdots , n\}$)  representing the CPU utilization observations of VMs hosted on a VMM}.
    
    \item \textbf{Output} \textit{ a subset of time ticks or a time tick where the behavior of the VMM is anomalous}.
    
\end{itemize}
\end{problem}

One of the significant challenges in this problem is the online detection, in which we receive the data incrementally, one time tick for each VM at a time, i.e., $X_{1}^j, X_{2}^j, \cdots$, for the $j^{th}$ VM. As we receive the data, the algorithm should output the time ticks where the behavior of the VMM  is observed as anomalous. However, without looking at the future few time ticks after time $t$, it would be impractical to determine whether at time point $t$, the VMM is anomalous or not since the time ticks ${t + 1, t + 2, \cdots}$, are essential in deciding whether an apparent detection at time $t$ was an actual or simply noise. Hence, we introduce a window parameter $w$, upon receiving a time tick $t + w$, the algorithm outputs whether at time $t$ the VMM showcased anomalous behavior or not. Additionally, as the change points for VMs hosted on VMM could be spread over a specific duration due to the effect of the actual fault being propagating to the VMs and the granularity of the collected monitoring data, therefore, using an appropriate window size can provide a way for getting those change points.


    
    

\subsection{Illustrative Example}
\begin{figure}[t]
\centerline{\includegraphics[width=0.55\linewidth]{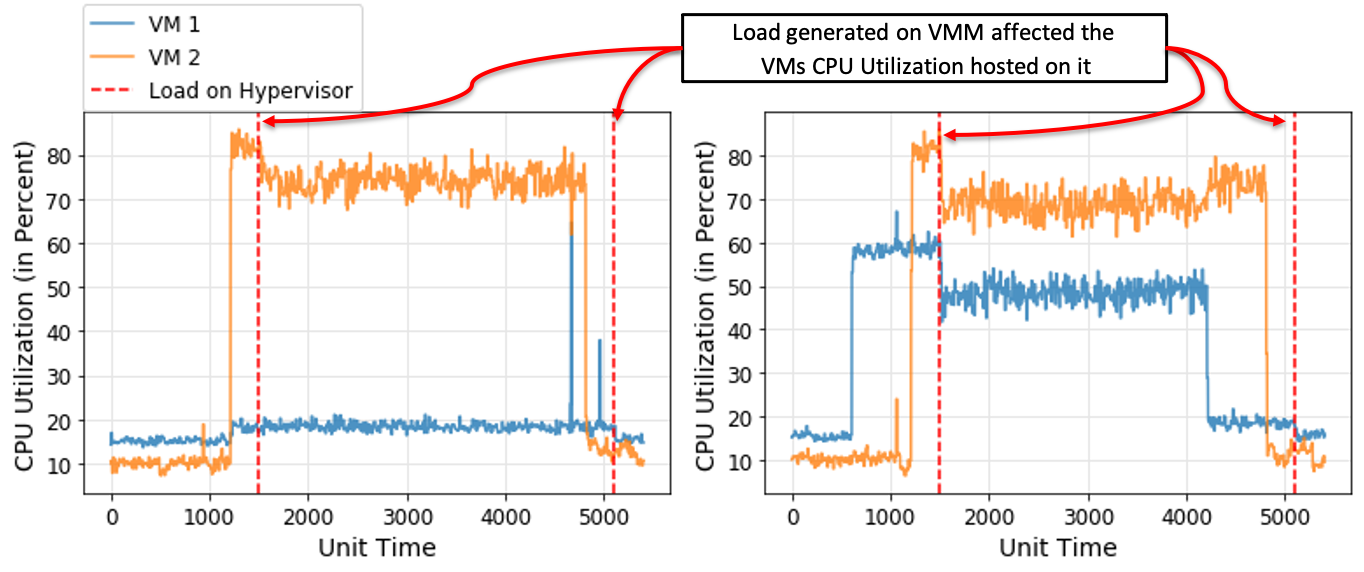}}
\caption{Examples showing CPU utilization of two virtual machines hosted on a VMM. The left sub-figure shows an application running only on VM 2, while the right sub-figure shows the application running on both VMs. We can see a significant decrement in the CPU utilization of the two VMs when an anomaly (high-CPU load) is generated on the VMM (shown by dotted red lines).}

\label{fig_example_problem}
\end{figure}
Here we illustrate the problem with two examples in Fig.~\ref{fig_example_problem} showcasing the CPU utilization of two virtual machines hosted on a VMM. In the left sub-figure, an application is running only on VM 2, while in the right, an application is running on both VMs. During the application run time, an anomaly, i.e., high CPU load, was generated on the hypervisor for some time (shown by dotted red lines). During this time, we can observe a significant drop in the CPU utilization by the application (affecting the performance of the application) of the two VMs (especially when an application is running on the VM). The load on a VMM affects all or most of the VMs hosted on it, which ultimately can significantly affect the performance of the applications running on the two VMs; therefore, we call such a VMM anomalous when the load was generated on it. 

\section{Indirect Anomaly Detection (IAD) Algorithm }
\label{sec:iad_algorithm}
This section presents our proposed Indirect Anomaly Detection (IAD) algorithm along with the implemented system for evaluating it. The overall system workflow diagram is shown in Figure~\ref{fig_iad_workflow} and mainly consists of two parts: the main \textit{IAD Algorithm}, and the \textit{Test Module} for evaluating the algorithm.

\begin{figure}[t]
\centerline{\includegraphics[width=0.8\linewidth]{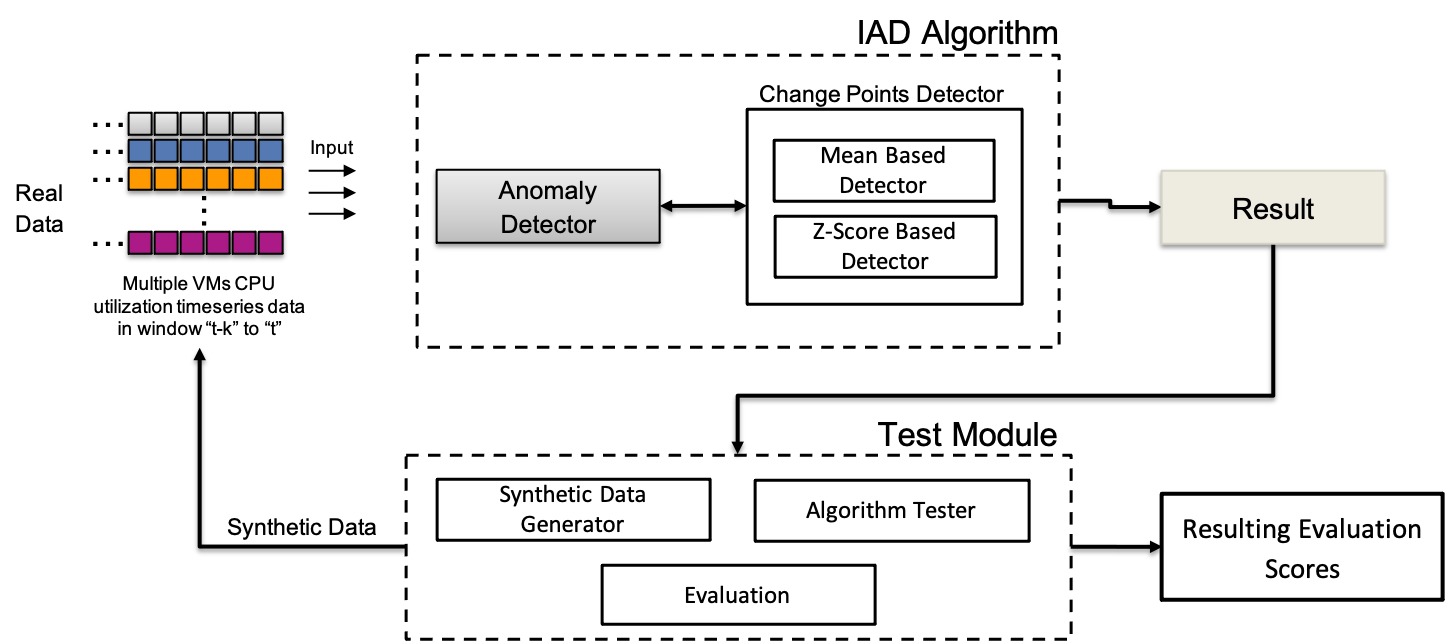}}
\caption{High-level system workflow of the implemented system for evaluating IAD algorithm and the interaction between its components in a general use case.}
\label{fig_iad_overall_workflow}
\end{figure}

\subsection{IAD Algorithm}
Our principal intuition behind the algorithm is that if a time tick $t$ represents a change point for some resource utilization (such as CPU utilization) in most VMs hosted on a VMM; then the VMM is also anomalous at that time tick. This is based on the fact that a fault in VMM will affect most of the VMs hosted on it, and therefore those VMs would observe a change point at a similar point of time (in the chosen window $w$ (Table~\ref{tab1:symbols})) in their resource's utilization.  IAD algorithm consists of two main parts, described below: 

\subsubsection{Change Points Detector}: We first explain how the change point, i.e., time tick where the time series changes significantly, is calculated. Recall from §\ref{sec:problem_statement} that, we have introduced a window parameter $w$, upon receiving the time tick $t+w$, the \textit{Change Points Detector} outputs whether the time tick $t$ is a change point or not. Given a dataset $X^j$ of size $w$ for $j^{th}$ VM, this component is responsible for finding the change points in that VM. This can be calculated in two ways: Mean-based detector and Z-score-based detector.

\begin{itemize}
    \item \textbf{Mean-based Detector}: In this detector, a $windowed\_mean$, i.e., the mean of all the values in the window, and the $global\_mean$, i.e., the mean of all the values until the current time tick is calculated. Since the IAD algorithm is designed for running it in an online way, therefore not all the values can be stored. Thus $global\_mean$ is calculated using Knuth’s algorithm~\cite{10.5555/270146, doi:10.1080/01621459.1974.10480219}. We then calculate the absolute percentage difference between the two means: $windowed\_mean$ and $global\_mean$. If the percentage difference is more significant than the specified threshold (by default is 5\%), then the time tick $t$ for $j^{th}$ VM is regarded as the change point.

\item  \textbf{Z-score-based Detector}: This detector is based on the calculation of the Z-scores~\cite{zscore, 9071555}. Similar to the Mean-based detector, here also a $windowed\_mean$, i.e., the mean of all the values in the window, and the $global\_mean$, i.e., the mean of all the values until the current time tick is calculated. We additionally calculate the $global\_stand\_deviation$, i.e., the standard deviation of all the values until the current time tick. Since the IAD algorithm is designed for running it in an online way, $global\_stand\_deviation$ is calculated using Welford's method~\cite{doi:10.1080/01621459.1974.10480219}. These statistics are then used for the calculation of the z-scores for all the data points in the window using Equation \ref{eq:1}.
\begin{equation}
z\_scores = \frac{(windowed\_mean - global\_mean)}{\frac{global\_stand\_deviation}{\sqrt{w}}}\label{eq:1}
\end{equation}
If the Z-scores of all windowed observations are greater than the defined threshold (3 $\times$ $global\_stand\_deviation$) then the time tick $t$ for $j^{th}$ VM is regarded as the change point.

\end{itemize}

In the main algorithm, only \textit{Z-Ssore-based Detector} is used as it provides higher accuracy and has fewer false positives. 

\subsubsection{Anomaly Detector}
\begin{figure}[t]
\centerline{\includegraphics[width=0.6\linewidth]{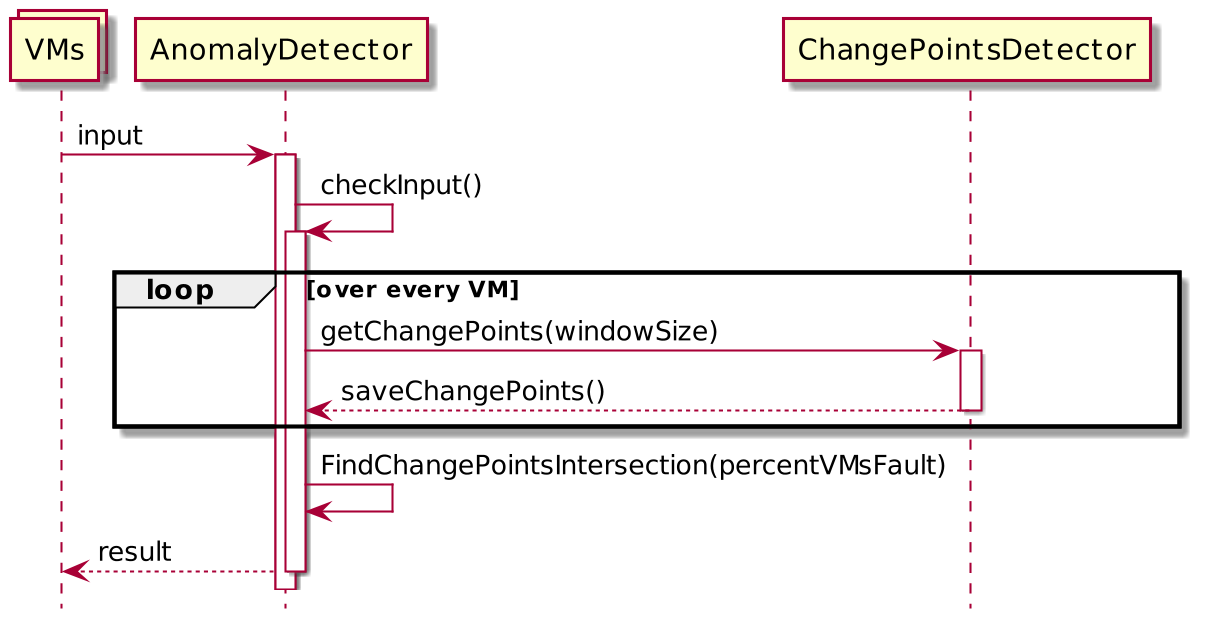}}
\caption{Indirect Anomaly Detection (IAD) Algorithm workflow sequence diagram}
\label{fig_iad_workflow}
\end{figure}

This component receives the input resource utilization data $X$ of size $n \times d$ where $d$ is the number VMs hosted on a VMM along with the \texttt{minPercentVMsFault} (Table~\ref{tab1:symbols})) as the input parameter.  We first check the input timeseries of $w$ length for 1) zero-length timeseries and 2) if the input timeseries of all VMs are of the same length or not. If any of the two initial checks are true, then we quit and don't proceed ahead. We assume that all the VM's resources utilization data is of the same length only.  After doing the initial checks, each of the VM's windowed timeseries belonging to the VMM is sent to the \textit{Change Points Detector} for the detection of whether the time tick $t$ is a change point or not. If the percentage number of VMs ($\{c_{t}^1, c_{t}^2, ..., c_{t}^m\}$ out of $d$) having the change point at time tick $t$ is greater than the \texttt{minPercentVMsFault} input parameter, then the VMM is reported as anomalous at time tick $t$. The above procedure is repeated for all time ticks. Figure~\ref{fig_iad_workflow} shows the workflow sequence diagram of the IAD algorithm. Furthermore, the developed approach can be applied for multiple VMMs as well.

\subsection{Test Module} This component is responsible for generating the synthetic data and evaluating the algorithm performance by calculating the F1-score on the results from the algorithm. It consists of multiple sub-component described below: 

\begin{itemize}
\item \textbf{Synthetic Data Generator}: It takes the number of VMMs, number of VMs per VMM, percentage of the VMs with a fault; as the input for generating synthetic timeseries data. This synthetic data follows a Gaussian distribution based on the input parameters. This component also automatically divides the generated data into true positive and true negative labels based on the percentage of the VMs with a fault parameter.

\item \textbf{Algorithm Tester}:  It is responsible for invoking the algorithm with various parameters on the synthetic data and tune the algorithm's hyperparameters.

\item \textbf{Evaluation}: The results from the algorithm are passed as the input to this sub-component, where the results are compared with the actual labels, and the overall algorithm score in terms of F1-score is reported. 

\end{itemize}

\section{Experimental Settings}
\label{sec:exp_settings}
We design our experiments to answer the following questions:

\textbf{Q1. Indirect Anomaly Detection Accuracy}: how accurate is IAD in the detection of anomalous VMM when compared to other popular algorithms?

\textbf{Q2. Anomalous VMMs finding efficiency and scalability}: How does the algorithm scale with the increase in the data points and number of VMs?

\subsection{Datasets}
\label{sec:evaluated_datasets}
For evaluating the IAD algorithm, we considered four types of datasets listed in Table~\ref{tab1:datasets} along with their information and are described below:

\textbf{Synthetic:} This is the artificially generated dataset using the \textit{Test Module} component described in \S\ref{sec:iad_algorithm}. 
    
\textbf{Experimental-Synthetic Merged:} This is a dataset with a combination of experimental data and synthetic data. We created two nested virtual machines on a VM in the Google Cloud Platform to collect the experimental dataset. 
The underneath VM instance type is n1-standard-4 with four vCPUs and 15 GB of memory, and Ubuntu 18.04 OS was installed on it. This VM instance acts as a host for the above VMs. \textit{libvirt} toolkit is used to manage and create nested virtualization on top of the host machine. Kernel-based Virtual Machine (KVM) is used as a VMM. The configuration of the two nested VMs are i) 2vCPU and 2GB memory, ii) 1vCPU and 1GB memory. Cloud-native web applications were run on these two VMs. Monitoring data from the two VMs and underneath host is exported using the Prometheus agent deployed on each of them to an external virtual machine. \textit{stress-ng} is used for generating the load on the VMM. Based on this infrastructure, we collected a dataset for various scenarios and combined it with the synthetic data. 

\textbf{Azure Dataset:} This dataset is based on the publicly available cloud traces data from Azure~\cite{10.1145/3132747.3132772}. We used the VMs data from it and created random groups of VMs, with each group representing the VMs hosted on a VMM. Afterward, we feed these timeseries groups in our synthetic data generator for randomly increasing or decreasing the CPU utilization of the VMs within a VMM based on the input parameters to create anomalous and non-anomalous VMMs. 

\textbf{Alibaba Dataset:} This dataset is based on the publicly available cloud traces and metrics data from Alibaba cloud~\cite{10.5555/3291168.3291175}. A similar method as the  \textit{Azure Dataset} was also applied to form this dataset. 

\begin{figure*}[t] 
\begin{subfigure}{.245\textwidth}
  \centering
  \includegraphics[width=1\linewidth]{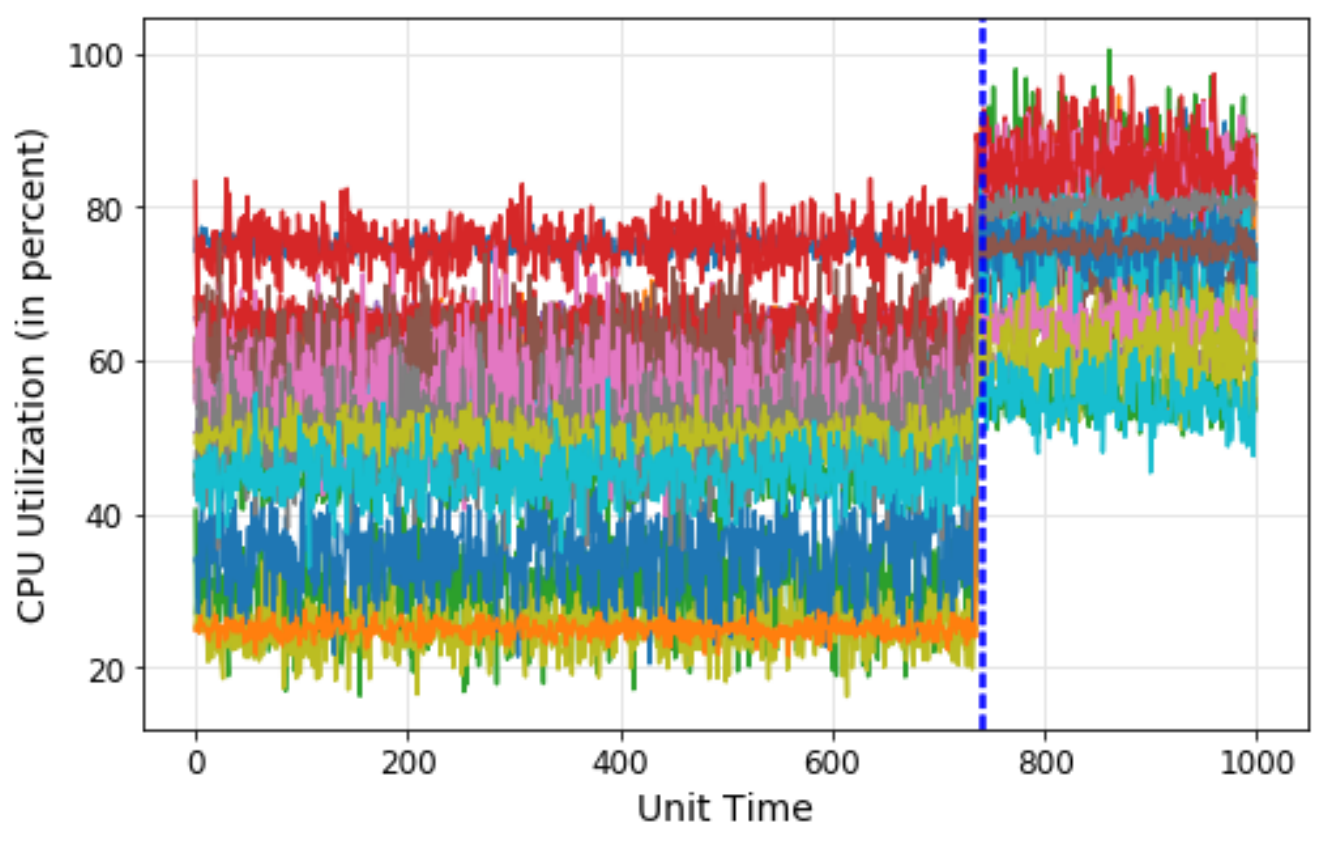}
    \captionof{figure}{Synthetic}
  \label{fig:synthetic_dataset}
\end{subfigure}%
\begin{subfigure}{0.245\textwidth}
  \centering
  \includegraphics[width=1\linewidth]{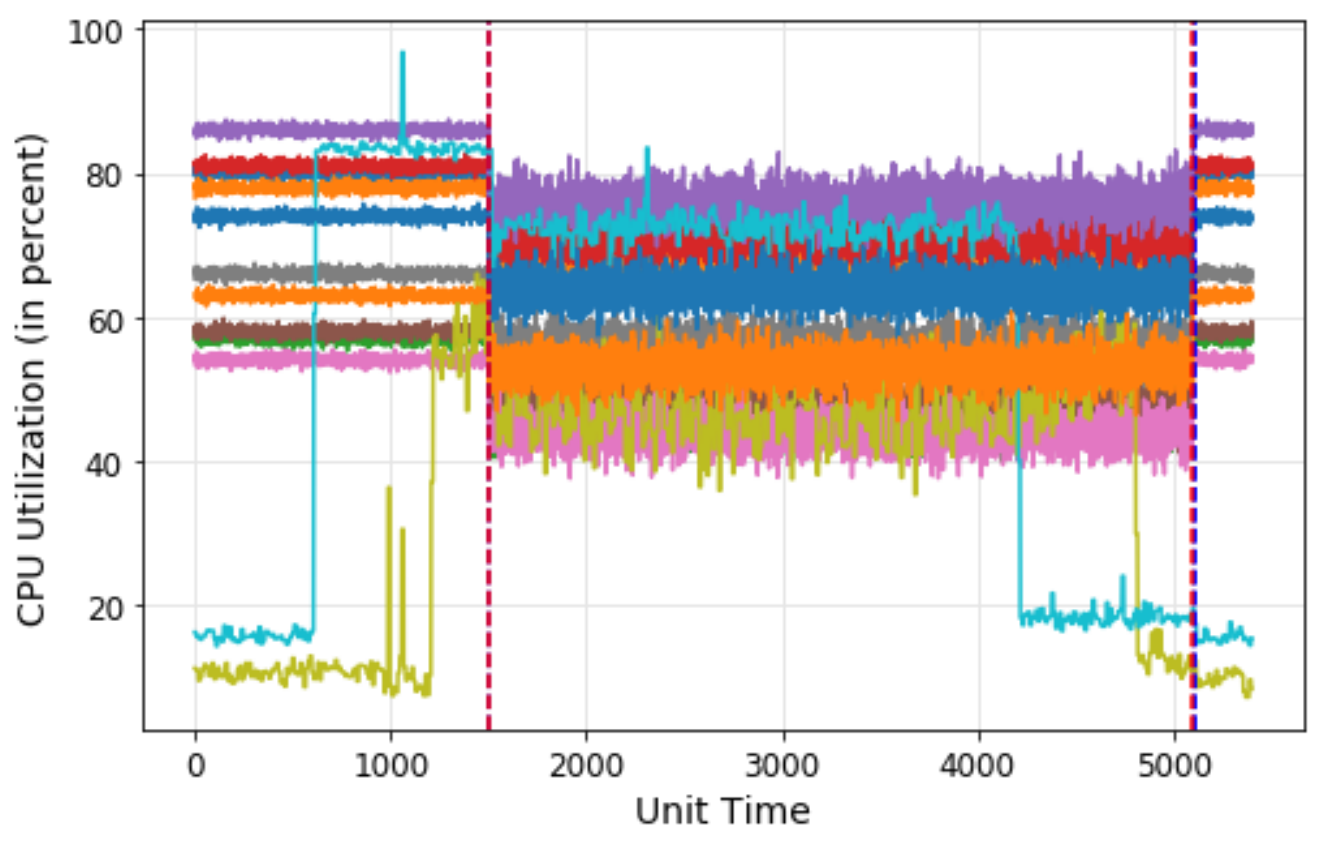}
      \captionof{figure}{Exp-Synthetic}
  \label{fig:exp_synthetic_dataset} 
\end{subfigure}
\begin{subfigure}{0.245\textwidth}
  \centering
  \includegraphics[width=1\linewidth]{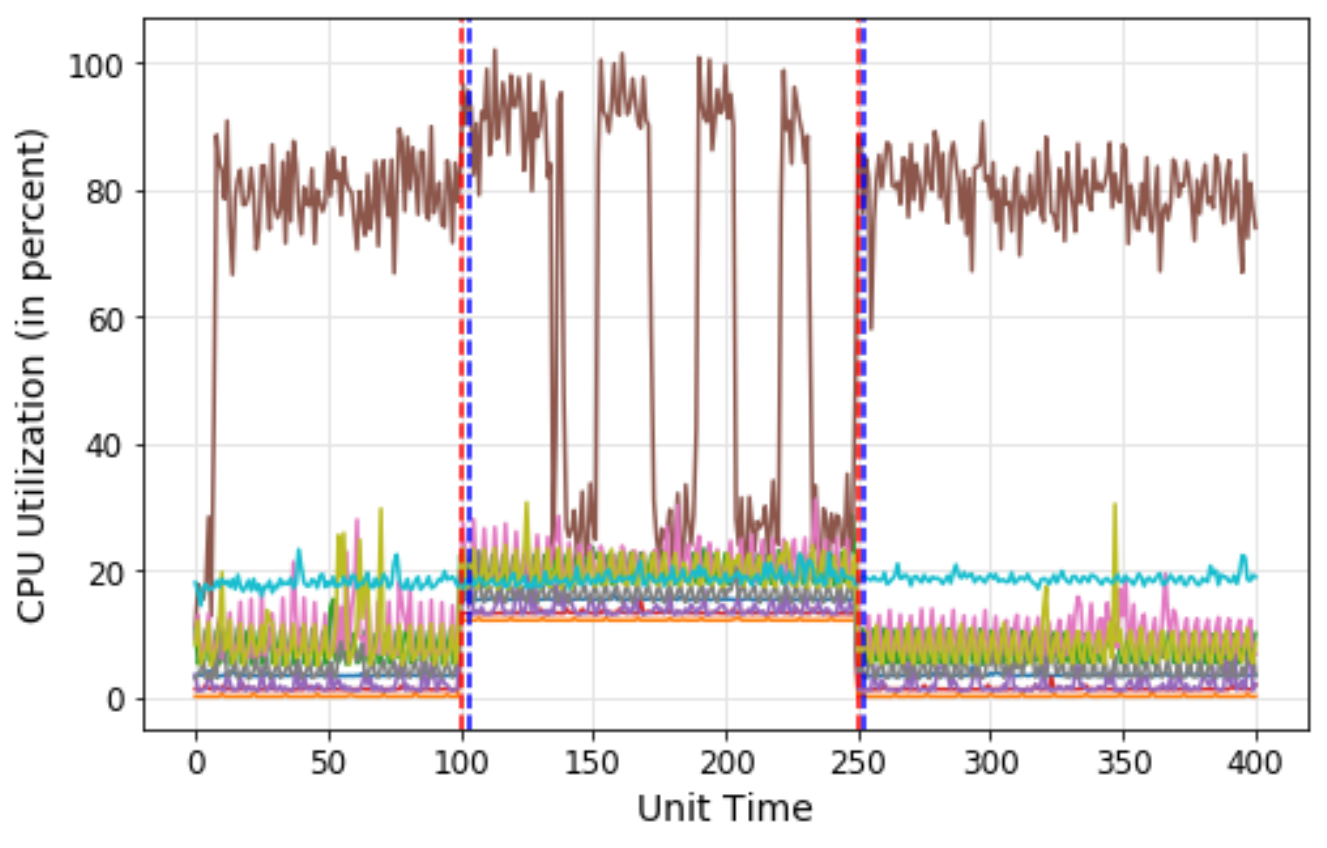}
      \captionof{figure}{Azure}
  \label{fig:azure_dataset} 
\end{subfigure}
\begin{subfigure}{0.245\textwidth}
  \centering
  \includegraphics[width=1\linewidth]{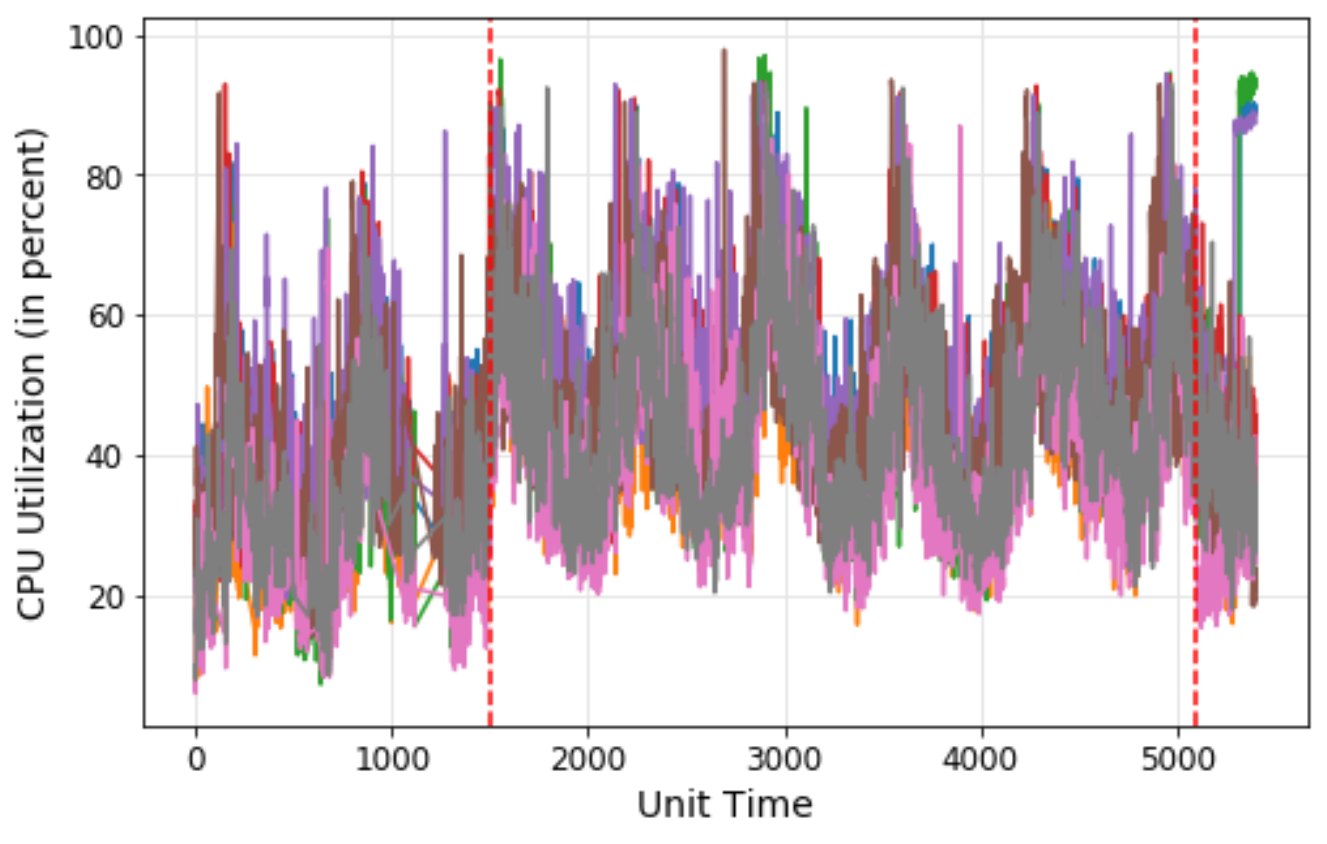}
      \captionof{figure}{Alibaba}
  \label{fig:alibaba_dataset} 
\end{subfigure}

 \caption{An example profile of an anomalous VMM having 10 VMs in all the datasets used in this work for evaluation.}
  \label{datasets_examples}
\end{figure*}

Figure~\ref{datasets_examples} shows an example profile of an anomalous VMM for all the datasets. 
\begin{table}[t]
\centering
\caption{Datasets used in this work for evaluating the algorithms.}\label{tab1:datasets}
\begin{tabular}{|c|c|c|c|c|c|}
\hline
\textbf{Dataset} &  \textbf{Anomalous}&  \textbf{Non-Anomalous}&  \textbf{VMs} &\textbf{TimeTicks} \\
\textbf{Name} &  \textbf{VMMs}&  \textbf{VMMs}&  \textbf{Per VMM} &\textbf{per VM} \\

\hline
Synthetic & 5 & 5 & 10 & 1000 \\
Exp-Synthetic Merged  & 42 & 17 & 2 (experimental) & 5400\\
&  &  & 8 (synthetic) & \\
Azure\textsuperscript{$\dagger$}~\cite{10.1145/3132747.3132772} & 16 & 10 & 10 & 5400 \\
Alibaba\textsuperscript{$\dagger$}~\cite{10.5555/3291168.3291175} & 10 & 10 & 10 & 5400 \\
\hline
\end{tabular}
{\raggedright \textsuperscript{$\dagger$}These are modified for our usecase. \par}
\end{table}

\subsection{Evaluated Algorithms}
\label{sec:algos_evaluated}
We compare IAD to the five other algorithms listed in Table~\ref{tab1:algos_used} along with their input dimension and parameters.  ECP is a non-parametric-based change detection algorithm that uses the E-statistic, a non-parametric goodness-of-fit statistic, with hierarchical division and dynamic programming for finding them~\cite{james2013ecp}. BnB (Branch and Border) and its online version (BnBO) are also non-parametric change detection methods that can detect multiple changes in multivariate data by separating points before and after the change using an ensemble of random partitions~\cite{Hooi2019BranchAB}. Lastly, we use the popular anomaly detection algorithm: isolation forest for detecting anomalous VMM~\cite{4781136}. The primary isolation forest (IF) works on the input data directly, while we also created a modified version of it called the isolation forest features (IFF), which first calculates several features such as mean, standard deviation, etc., for all values within a window on the input dataset and then apply isolation forest on it. The downside of the IF and IFF is that they require training.

\subsection{Other Settings}
We have used F1-Score (denoted as F1) to evaluate the performance of the algorithm. Evaluation tests have been executed on 2.6 GHz 6-Core Intel Core i7 MacBook Pro, 32 GB RAM running macOS BigSur version 11. We implement our method in Python. For our experiments, hyper-parameters are set as follows. The window size $w$ is set as 1 minute (60 samples, with sampling done per second), threshold $k$ as 5\%, and percentVMsFault $f$ as 90\%. However, we also show experiments on parameter sensitivity in this section.

\begin{table}[t]
\centering
\caption{The details of the algorithms used in this work for evaluation, along with their input dimension and parameters.}\label{tab1:algos_used}
\begin{tabular}{|l|c|l|}
\hline
\textbf{Algorithm} &  \textbf{Input Dimension}&  \textbf{Parameters}\\
\hline\hline
IAD & n $\times$ d & w, minPercentVMsFault \\
ECP~\cite{james2013ecp}  & n $\times$ d  & change points, Min. points b/w change points
 \\
BNB~\cite{Hooi2019BranchAB}  & n $\times$ d  & w, number of trees, threshold for change points
 \\
BNBOnline~\cite{Hooi2019BranchAB}  & n $\times$ d  & w, number of trees, threshold for change points
 \\
IF~\cite{4781136} & n $\times$ d & contamination factor \textcolor{red}{(requires training)}
 \\
IFF~\cite{4781136} & n $\times$ features  & contamination factor \textcolor{red}{(requires training)}
 \\

\hline
\end{tabular}
\end{table}

\section{Results}
\label{sec:results}
Our Initial experiments showed that 1) CPU metric is the most affected and visualized parameters in the VMs when some load is generated on the VMM; 2) All or most VMs are affected when a load is introduced on the VMM. 
\subsection{Q1. Indirect Anomaly Detection Accuracy}
\label{sec:est_time_accuracy}
Table~\ref{tab1:accuracy_score} shows the best F1-score corresponding to each algorithm evaluated in this work (\S\ref{sec:algos_evaluated}) and on all the datasets (\S\ref{sec:evaluated_datasets}). We can observe that \textit{IAD} algorithm outperforms the others on two datasets, except for the Experiment-Synthetic dataset (BNB performed best with F1-Score of \texttt{0.90}) and Alibaba dataset (IFF performed best with F1-Score of \texttt{0.66}. However, if one wants to find an algorithm that is performing well on all the datasets (Average F1-score column in Table~\ref{tab1:accuracy_score}), in that case, \textit{IAD} algorithm outperforms all the others with an average F1-score of \texttt{0.837} across all datasets. 

\begin{table}[t]
\centering
\caption{F1-score corresponding to each algorithm evaluated in this work (\S\ref{sec:algos_evaluated}) and on all the datasets (\S\ref{sec:evaluated_datasets})}\label{tab1:accuracy_score}
\begin{tabular}{|l|c|c|c|c|c|c|}
\hline
\textbf{Algorithm} &  \textbf{Synthetic}&  \textbf{Exp-Synthetic}& \textbf{Azure}& \textbf{Alibaba} &  \textbf{Average F1-score}\\
\hline\hline
IAD & \textbf{0.96} & 0.86 & \textbf{0.96} &0.57 & \textbf{0.837} \\
ECP  & 0.67 & - & 0.76 &0.51  &0.64\\
BNB  & 0.62 & \textbf{0.90} & 0.8 &0.33  &0.662\\
BNBOnline  & 0.87 & 0.81 & 0.86 &0.4  &0.735\\
IF & 0.76 & 0.83 & 0.76 & 0.2  &0.637\\
IF Features (IFF) & 0.76 & 0.83 & 0.76 & \textbf{0.66} &0.75 \\

\hline
\end{tabular}
\end{table}

\begin{figure*}[t] 
\centering
\begin{subfigure}{.35\textwidth}
  \centering
  \includegraphics[width=1\linewidth]{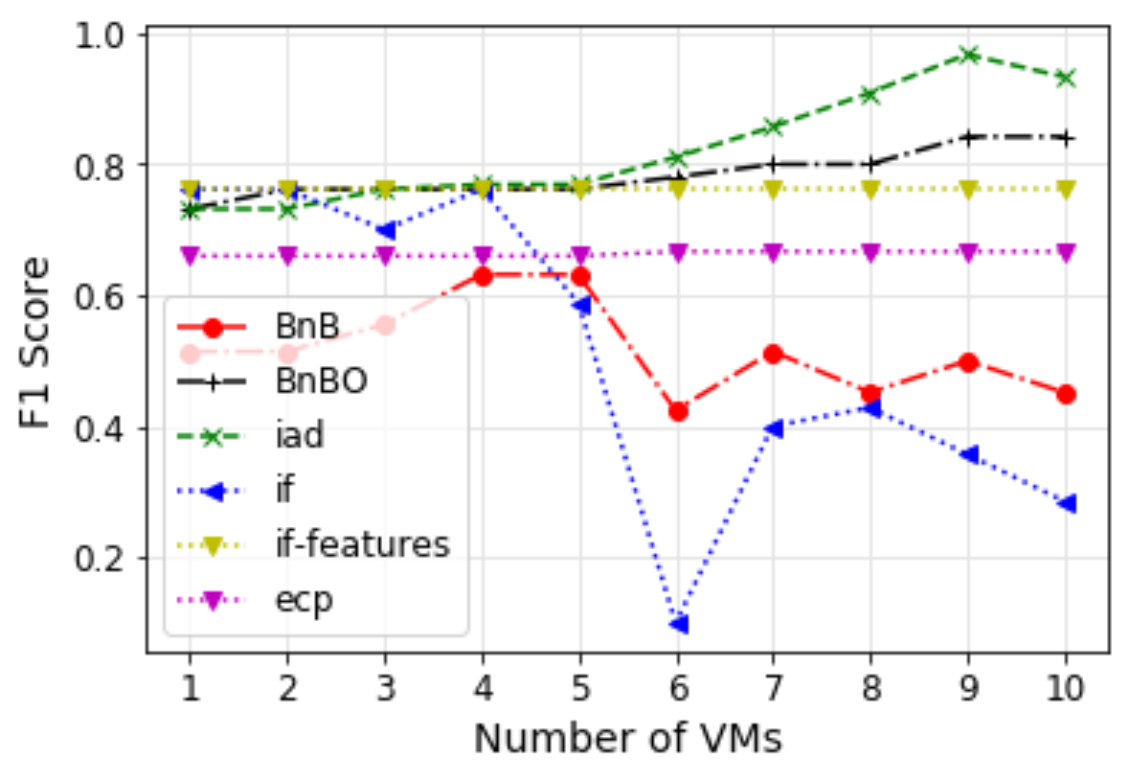}
    \captionof{figure}{Synthetic}
  \label{fig:synthetic_f1}
\end{subfigure}%
\begin{subfigure}{0.35\textwidth}
  \centering
  \includegraphics[width=1\linewidth]{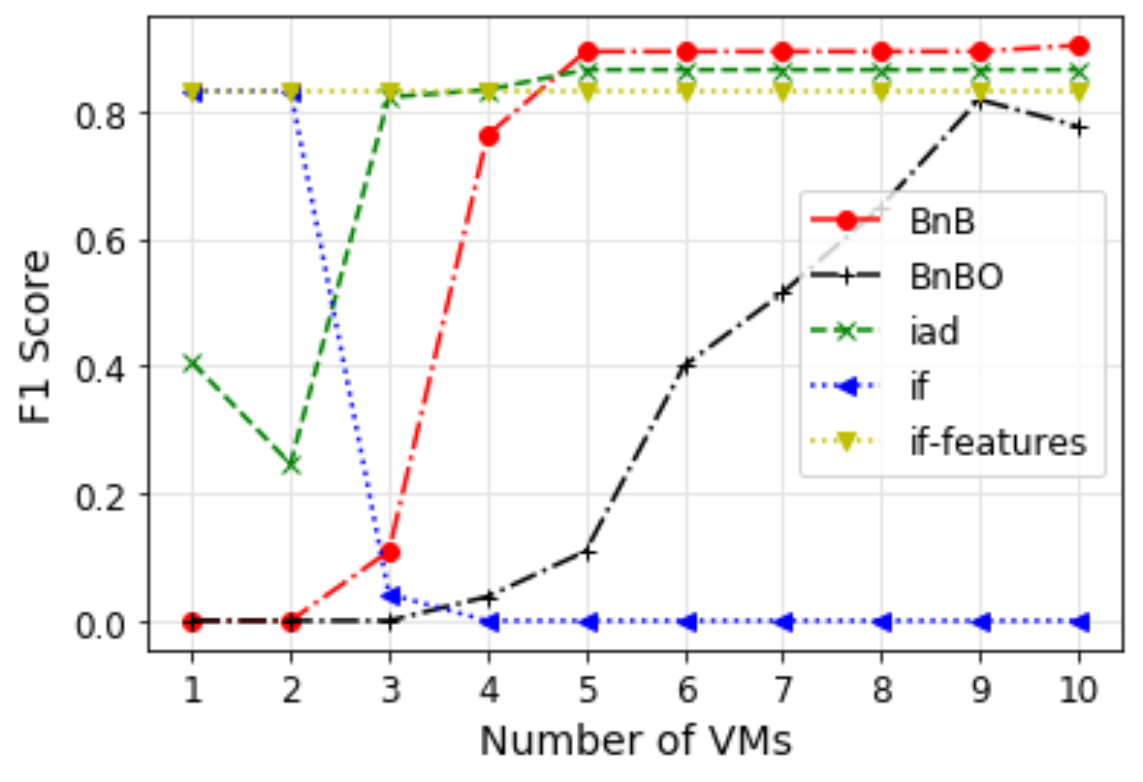}
      \captionof{figure}{Exp-Synthetic}
  \label{fig:exp_synthetic_f1} 
\end{subfigure}
\begin{subfigure}{0.35\textwidth}
  \centering
  \includegraphics[width=1\linewidth]{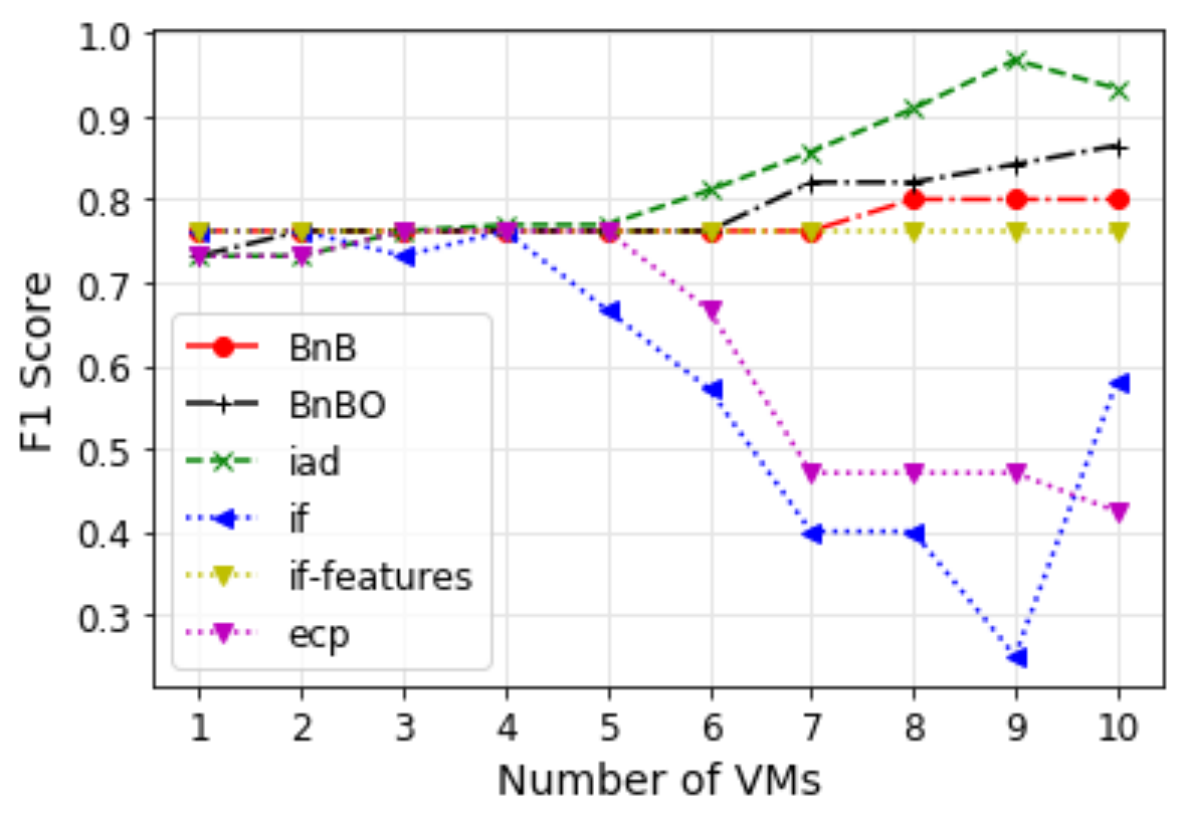}
      \captionof{figure}{Azure}
  \label{fig:azure_f1} 
\end{subfigure}
\begin{subfigure}{0.35\textwidth}
  \centering
  \includegraphics[width=1\linewidth]{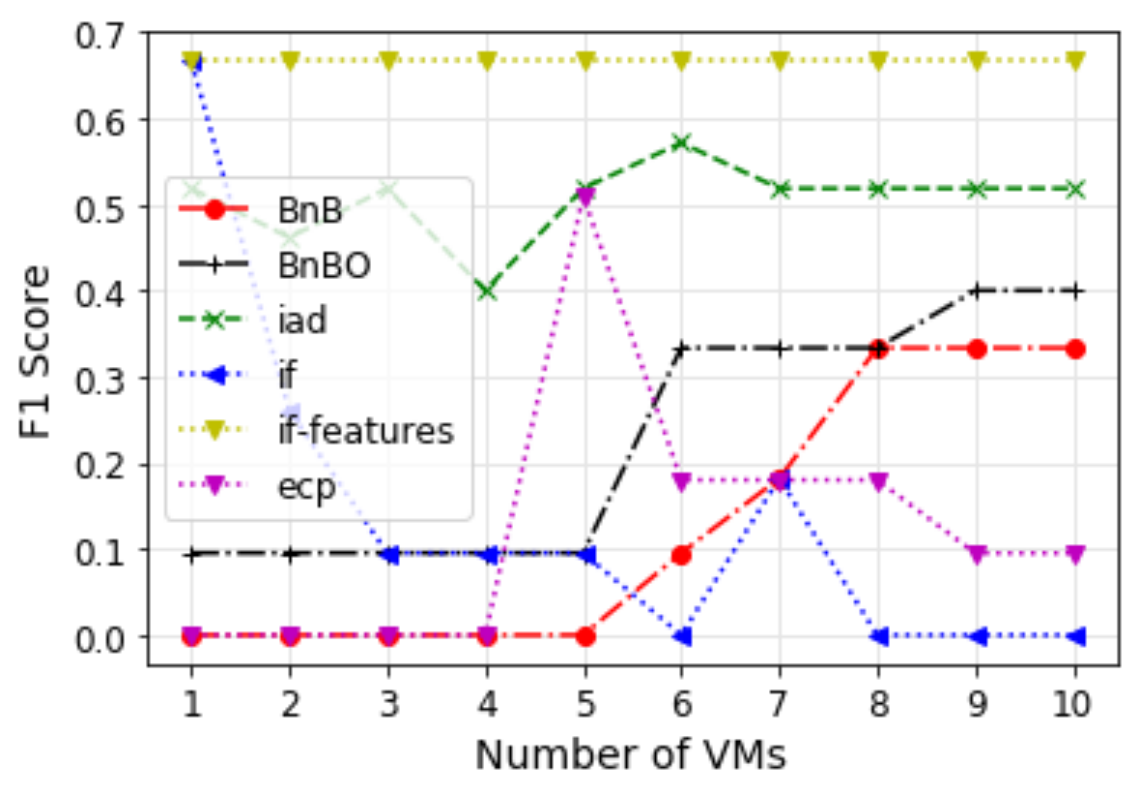}
      \captionof{figure}{Alibaba}
  \label{fig:alibaba_f1} 
\end{subfigure}

 \caption{F1-score variation with the number of VMs corresponding to each algorithm evaluated in this work (\S\ref{sec:algos_evaluated}) and on all the datasets (\S\ref{sec:evaluated_datasets})}
  \label{algorithms_f1_results}
\end{figure*}

Furthermore, we present the detailed results of the algorithms on all four datasets varying with the number of VMs and are shown in Figure~\ref{algorithms_f1_results}. One can observe that \textit{IAD} performs best across all the datasets, and its accuracy increases with the increase in the number of VMs. Additionally, after a certain number of VMs, the F1-score of \textit{IAD} becomes stable. This shows that if, for example, we have the synthetic dataset, then the best performance is possible with VMs $\geq$ \texttt{9}. Similarly, in the case of the Azure dataset, while for the Exp-Synthetic dataset, one needs at least five VMs, and for the Alibaba dataset, seven VMs for the algorithm to perform well.

\subsection{ Q2. Anomalous VMMs finding efficiency and scalability}
\label{sec:config_finding_efficency_scalability}
\begin{figure*}[t] 
\centering
\begin{subfigure}{.35\linewidth}
  \centering
  \includegraphics[width=1\linewidth]{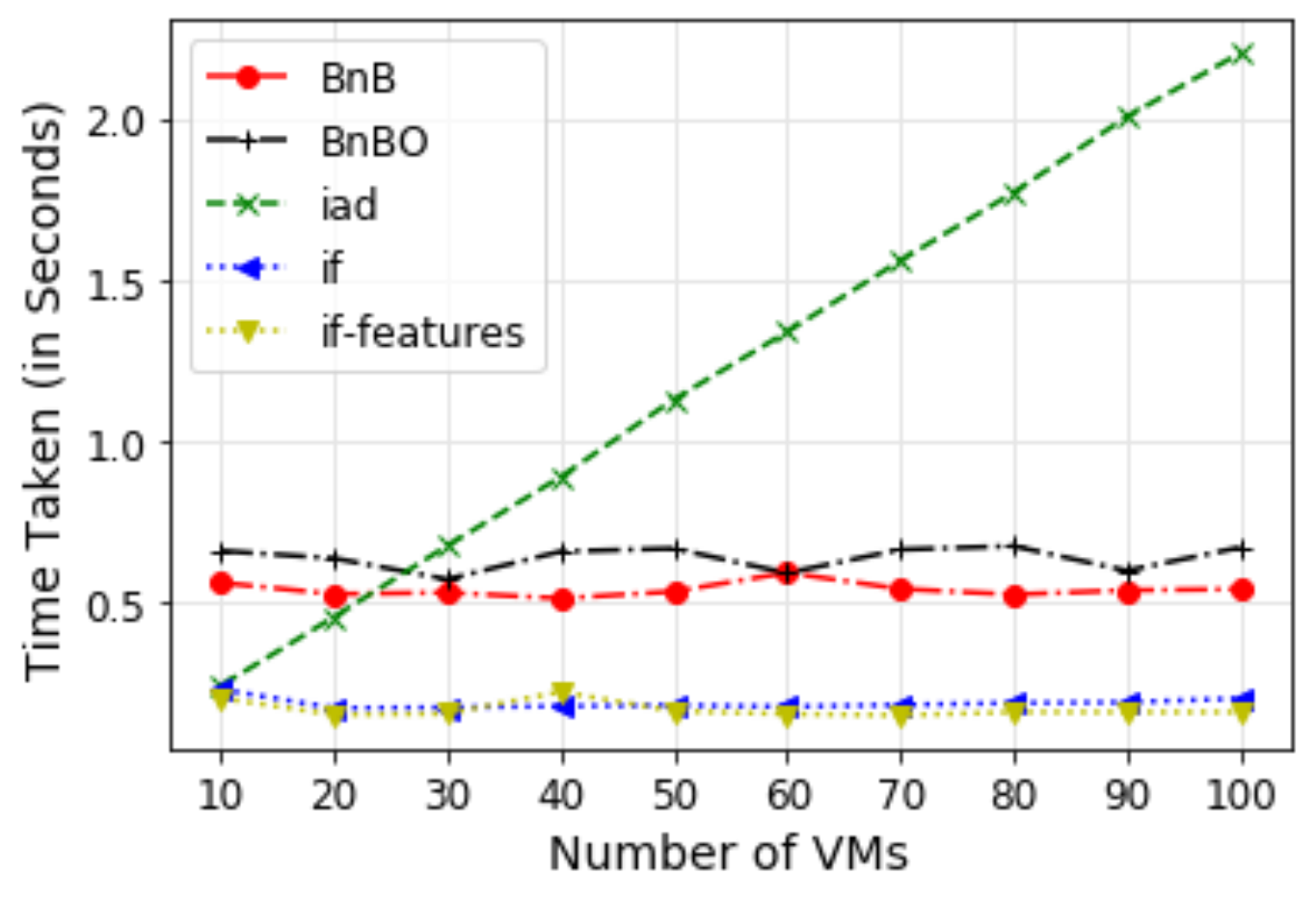}
    \captionof{figure}{With number of VMs}
  \label{fig:train_time}
\end{subfigure}
\begin{subfigure}{0.35\linewidth}
  \centering
  \includegraphics[width=1\linewidth]{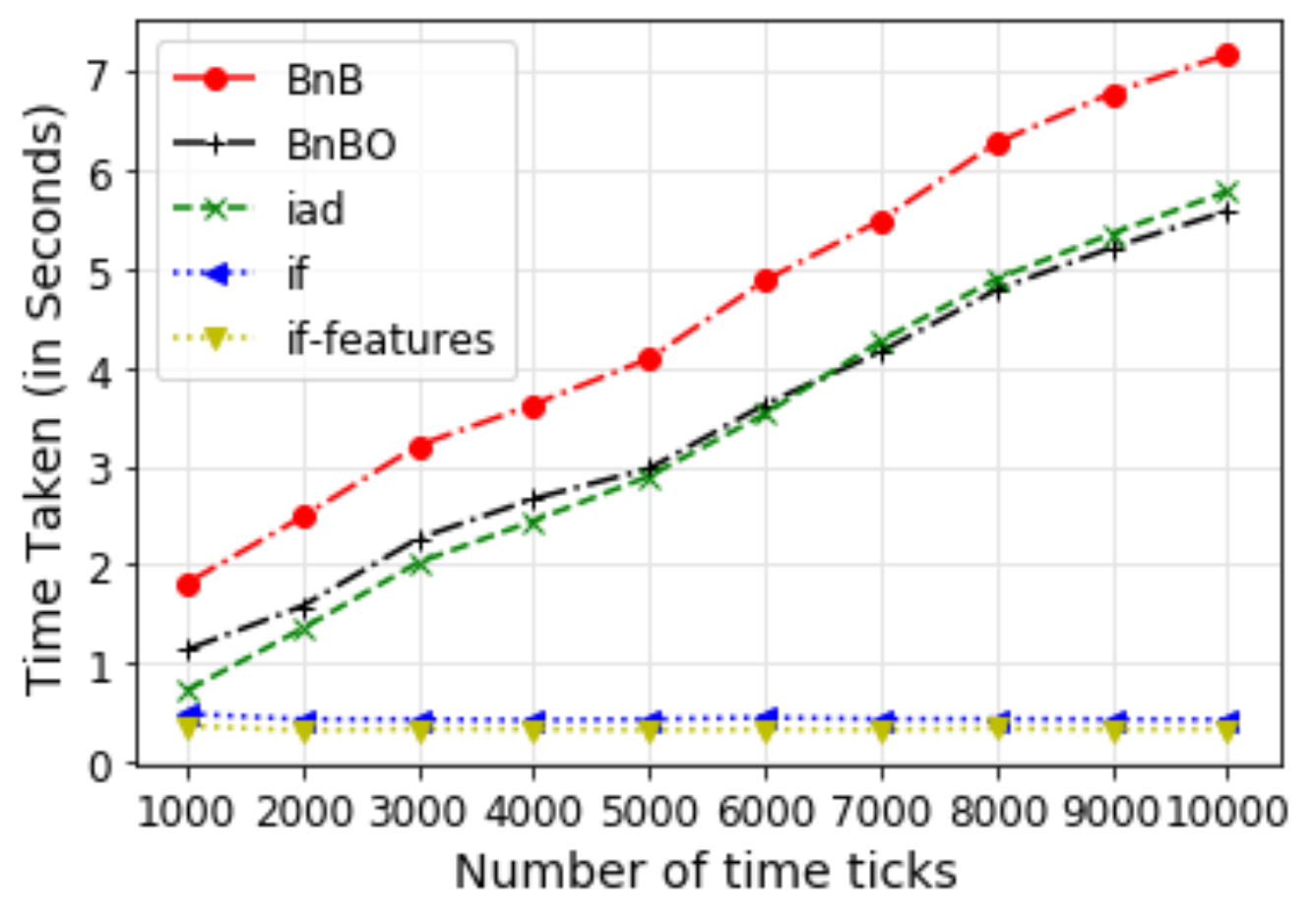}
      \captionof{figure}{With number of time ticks }
  \label{fig:predict_time} 
\end{subfigure}
 \caption{Algorithm's detection method scalability with respect to different parameters.}
  \label{algorithms_time_results_scalability}
\end{figure*}

Next, we verify that our algorithm's detection method scale linearly and compare it against other algorithms. This experiment is performed with the synthetic dataset, since we can increase the number of VMs per VMM in it. We linearly increased the number of VMs from 1 to 100 and repeatedly duplicated our dataset in time ticks by adding Gaussian noise. Figure~\ref{algorithms_time_results_scalability} shows various algorithm's detection method scalability for different parameters. One can observe that \textit{IAD's} detection method scale linearly in terms of both the parameters. However, when the number of VMs are scaled to \texttt{100}, \textit{IAD} takes a longer time as compared to others, but it provides results under \texttt{2.5s} which if we see is not that much considering the accuracy we get with that algorithm.   However, on the time ticks parameter, \textit{BNB}, \textit{BNBOnline} and \textit{IAD} performed similar to each other, while \textit{IF} and \textit{IFF} provides results under \texttt{1} second, but its accuracy is worse as compared to the others on all the datasets, and it has the extra overhead of training. \textit{ECP} algorithm's results are not shown, since it requires more than an hour for performing the detection with \texttt{100} VMs and \texttt{100,000} time ticks. 


\section{Conclusion}
\label{sec:conclusion}
We propose \textit{IAD} algorithm for indirect detection of anomalous VMMs by solely using the resource's utilization data of the VM's hosted on them as the primary metric. We compared it against the popular change detection algorithms, which could also be applied to the problem. We showcased that \textit{IAD} algorithm outperforms all the others on an average across four datasets by \texttt{11\%} with an average accuracy score of \texttt{83.7\%}. We further showcased that \textit{IAD} algorithm scale's linear with the number of VMs hosted on a VMM and number of time ticks. It takes less than \texttt{2.5} seconds for \textit{IAD} algorithm to analyze 100 VMs hosted on a VMM for detecting if that VMM is anomalous or not. This allows it to be easily usable in the cloud environment where the fault-detection time requirement is low and can quickly help DevOps to know the problem is of the hypervisor or not. 

The future direction includes using other metrics like network and storage utilization to enhance the algorithm's accuracy further.


\bibliographystyle{splncs04}
\bibliography{bib}
\end{document}